\documentclass[aps,prl,twocolumn,reprint,floatfix,showpacs,groupedaddress]{revtex4}
\usepackage{graphicx}
\usepackage{psfrag}
\usepackage{color}
\usepackage{amsbsy}
\usepackage{amssymb}

\def\a{\alpha}
\def\b{\beta}

\def\g{\gamma}
\def\dd{\mbox{d}}

\def\s{\sigma}

\begin{document}
\bibliographystyle{apsrev}


\title{Brownian ratchets driven by asymmetric nucleation of hydrolysis waves}



\author{Amit Lakhanpal and Tom Chou}
\affiliation{Dept. of Biomathematics \& Dept. of Mathematics, UCLA,
  Los Angeles, CA 90095-1766}


\date{\today}

\begin{abstract}
We propose a stochastic process wherein molecular transport is mediated
by asymmetric nucleation of domains on a one-dimensional substrate.
Track-driven mechanisms of molecular transport arise in biophysical
applications such as Holliday junction positioning and collagenase
processivity. In contrast to molecular motors that hydrolyze
nucleotide triphosphates and undergo a local molecular conformational change, we
show that asymmetric nucleation of hydrolysis waves on a track can
also result in directed motion of an attached particle. 
Asymmetrically cooperative kinetics between ``hydrolyzed''
and ``unhydrolyzed'' states on each lattice site generate
moving domain walls that push a particle sitting on the track. We use
a novel fluctuating-frame, finite-segment mean field theory to
accurately compute steady-state velocities of the driven particle and
to discover parameter regimes which yield maximal domain wall flux,
leading to optimal particle drift.
\end{abstract}

\maketitle


Molecular motors such as kinesins, myosins, helicases, and polymerases
typically convert part of the free energy of ATP or GTP hydrolysis to
a conformational change \cite{MOTOR}.  This molecular deformation
leads to motion of the motor against a load on a track.
Although the literature on such molecular motors is vast, much less
attention has been paid to the theory of molecular motions that
exploit the dynamics of the track on which translation occurs. Such
loads are propelled by the track, which itself is undergoing catalyzed
state changes by, {\it e.g.}, hydrolysis.

Two biological strategies involving track-propelled particles are
collagenase catalysis and Holliday junction transport.
Collagenase MMP-1, an enzyme that associates with and cleaves
collagen, is propelled by proteolysis of the collagen track
\cite{COLL,SAFF}.
The cleaving of bonds prevents the collagenase from diffusing back
across the broken bond, resulting in biased transport of the
collagenase. Thus the statistical dynamics of the track propels the
enzyme.  This dynamic has been modeled by a burnt bridge model
\cite{BB0,BB1,SAFF,KOLO}.

Another system where substrate modification possibly leads to biased
motion is the translocation of Holliday junctions \cite{KLAP}.  The Holliday
junction at which two double-stranded DNA molecules exchange one of
their strands may be moved by the dynamics of hydrolysis states of the
DNA binding protein RecA. RecA polymerizes on at least one of the
strand-exchanging dsDNA molecules, assembling into a long nucleoprotein
filament. The RecA monomers appear to hydrolyze ATP and can exist in
different states, much like the intermediate hydrolysis states of
myosin motors. The dynamics of the interconversion among these
hydrolysis states may provide the force necessary to rotate DNA
strands about each other during Holliday junction translocation. An
especially promising model of this process exploits asymmetric
cooperativity in the hydrolysis of the nucleotide triphosphate
cofactors associated with each RecA monomer.  This intrinsic asymmetry
of the filament gives rise to ``waves'' of hydrolyzed monomers with a
preferred direction, thereby moving the junction by virtue of its
preferential attachment to the hydrolyzed segment of the RecA filament
\cite{KLAP}.  These examples constitute only two of many mechanisms
through which chemical energy may be harnessed to perform mechanical
work by the substrate rather than a motor protein. In this Letter, we
develop a general stochastic theory of track-driven, hydrolysis
wave-mediated transport.  In addition to analyzing our model using Monte-Carlo
(MC) simulations, we also formulate a moving-frame mean field theory
(MFT) that accurately predicts novel features of the transport.

\begin{figure}[htb]
\includegraphics[width=2.5in]{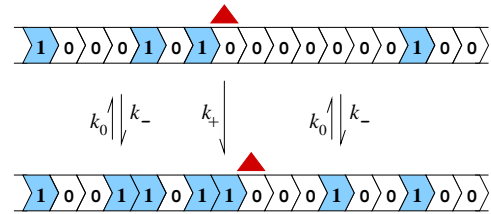}
\caption{Schematic of the asymmetric nucleation 
process. An intrinsic asymmetry in the lattice sites 
gives rise to asymmetric cooperativity and nucleation.
The transported particle is represented by a triangle.}
\label{FIG1}
\end{figure}

As in models of the ATP cycle of myosin motors, in RecA hydrolysis
wave-mediated transport the RecA subunits can exist in
a number of substates corresponding to sites that have bound ATP,
ADP+P$_{i}$, ADP, or are empty. To simplify our model, we will assume
that each site of the substrate lattice exists in only one of two
possible states, ``hydrolyzed'' ($\sigma = 1$) and ``unhydrolyzed''
($\sigma = 0$).  Any lattice site $i$ can transition from state
$\s_{i}=1$ to state $\s_{i}=0$ with rate $k_{0}$. The reverse process,
physically corresponding to ``hydrolysis'' or ``nucleation,'' fills an
empty site.  In our model, an asymmetry arises in the nucleation
transitions $\s_{i} = 0 \rightarrow \s_{i}=1$. If site $i-1$ is also
in the state $\s_{i-1}=0$, then the transition $\s_{i}=0 \rightarrow
\s_{i} = 1$ occurs with rate $k_{-}$.  However, if $\s_{i-1}=1$, then
the transition $\s_{i}=0 \rightarrow \s_{i} = 1$ occurs with rate
$k_{+}$.
If $k_{+} \neq k_{-}$, the process is asymmetric and can lead to a
steady-state current of domain walls. If a particle is associated with
the lattice, it will be pushed each time a domain wall passes
it. Thus, a net flux of domain walls will lead to directed particle
transport. The kinetics of the lattice is outlined in Fig. \ref{FIG1}.





The corresponding Master equation is similar to that which describes
Glauber dynamics of a one-dimensional Ising model, except that the
asymmetry in the transition rates prevents this system from supporting
an equilibrium state. Since no exact solutions are known, we employ a
hybrid finite-segment mean field theory (MFT) and MC simulations to
obtain quantitative results and physical understanding.


First consider a translationally invariant (infinite or periodic)
lattice in the absence of an associated load particle.  In the fixed
laboratory frame, the moments $\langle \s_{i}\s_{j}\ldots\rangle$ of
the hydrolysis states can be derived from the Master equation.  These
hierarchical moment equations are not closed.  For example, the
equation for the first moment

\begin{equation}
\begin{array}{l}
\displaystyle {\dd \langle \s_{i}\rangle \over \dd t} = k_{-}-(k_{-}+k_{0})\langle\s_{i}\rangle 
+ \Delta\langle\s_{i-1}\rangle - \Delta\langle\s_{i}\s_{i-1}\rangle,
\label{MOMENT1}
\end{array}
\end{equation}

\noindent where $\Delta \equiv k_{+}-k_{-}$ is the hydrolysis
asymmetry, depends on correlations $\langle\s_{i}\s_{i-1}\rangle$.
The simplest mean field approximation assumes
$\langle\s_{i}\s_{i-1}\rangle
=\langle\s_{i-1}\rangle\langle\s_{i}\rangle$, which, when combined
with the steady-state limit ($\dd \langle\s_{i}\rangle/\dd t = 0$) of
Eq. \ref{MOMENT1} gives a steady-state mean hydrolysis level

\begin{equation}
\langle\s_{i}\rangle = {k_{+}-2k_{-}-k_{0} \over 2\Delta} + 
{\sqrt{(k_{+}-k_{0})^{2}+4k_{-}k_{0}}\over 2\Delta}.
\label{MFT1}
\end{equation}

\begin{figure}[h]
\includegraphics[width=3.4in]{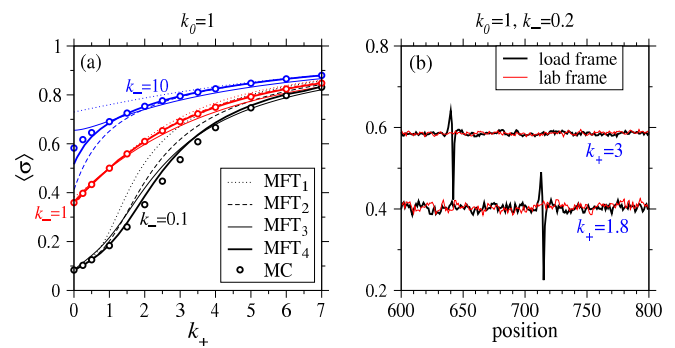}
\caption{(a) Mean densities computed from finite-segment MFT and from
MC simulation. The three groups of densities correspond to $k_{-} =
0.1,1,10$. Within each group, finite-segment mean field results with
sizes $m=1,2,3,4$ (denoted MFT$_{m}$) are compared with results from
MC simulations (open circles).  (b) Density profiles (for
$k_{+}=1.8,3$) in the lab frame (thin light curves) and in a frame
moving with the load particle (thick dark curves) that follows the
rules indicated in Fig. \ref{STEPS}.}
\label{FIG2}
\end{figure}

\noindent As shown by the dotted curves in Fig. \ref{FIG2}a,  this result is
only in qualitative agreement with the mean hydrolysis level obtained 
from MC simulations on a lattice with $N=1000$ sites(open circles).

The locality of the asymmetric interactions suggests that correlations
are short-ranged as in the totally asymmetric exclusion process
\cite{TASEP0,TASEP1}.  Thus, more accurate approximations can be
systematically implemented by considering small clusters in which all
possible configurations are identified, and enumerating all the
transitions among them.  The densities at both ends of this cluster are
then self-consistently matched to bulk values inferred by the
statistics within the cluster. This finite-segment mean field approach
has been used to study the nonequilibrium steady-states of related
models such as the asymmetric exclusion process \cite{DEFECT,NOWAK}.
For example, consider all possible configurations in a segment of
$m=2$ lattice sites.  If we enumerate the states corresponding to the
binary representation of the state number, ({\it i.e.}, $P_{0}=00,
P_{1} = 01, P_{2} = 10, P_{3}=11$), the $2^{2}\times 2^{2}$ transition
matrix defined by $\dot{{\bf P}} = {\bf M}{\bf P}$ is

\begin{equation}
{\bf M}  =\left[ \begin{array}{cccc}
-2k_{-}-s\Delta  & k_{0} &k_{0} & 0 \\[13pt]
k_{-} & -k_{0}-k_{-}-s \Delta & 0 & k_{0} \\[13pt]
k_{-}+s\Delta  & 0 & -k_{+}-k_{0} & k_{0} \\[13pt]
0 & k_{-}+s\Delta & k_{+} & -2k_{0} \end{array}\right],
\label{MATRIX}
\end{equation}

\noindent where ${\bf P} = (P_{0},P_{1},P_{2},P_{3})^{T}$ is the
probability vector and $s$ is the mean occupancy of the site
immediately to the left of the explicitly enumerated pair. Since $s$
represents the mean occupation of the rightmost site of the preceding
segment, we impose self-consistency by setting $\sum_{i=odd}P_{i} = s$
and solving for $s$ numerically. The simple mean field approximation
(Eq. \ref{MFT1}) corresponds to $m=1$.

Fig. \ref{FIG2}a shows the increasing accuracy in determining
$\langle\s\rangle$ upon using larger $m$ (shown are $m=1,2,3,4$) in
the finite-segment mean field approach. Although $m=1$ (simple mean
field theory) can give results appreciably disparate from MC
simulation results, larger clusters ($m=2,3,4, \ldots$) significantly
improve convergence to the correct mean density level.
Moreover, simple MFT (Eq. \ref{MFT1}) is exact in the
symmetric, equilibrium limit $k_{+} \rightarrow k_{-}$, where the
moment equations are closed.

\begin{figure}
\includegraphics[width=3.3in]{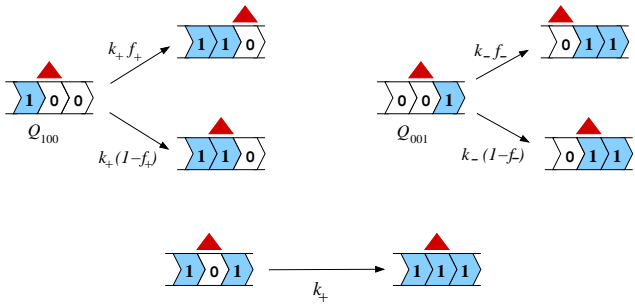}
\caption{The kinetic rules for domain wall-driven particle motion.  A
particle is pushed forward with probability $f_{+}$ whenever a $\ldots
100 \ldots$ domain wall tries to move past it from left to
right. Similarly, the particle moves backward with rate $k_{-}f_{-}$}
\label{STEPS}
\end{figure}

Now consider a lattice-associated particle that can be moved by the
nonequilibrium fluctuations inherent in the substrate. Simple kinetic
rules are defined in Fig. \ref{STEPS}.
Without loss of generality, we also assume that the particle does not
have intrinsic fluctuations. Since its stochastic motion is caused
only by the asymmetric domain walls fluctuating past it, we must
determine the domain wall probability {\it in the frame of the moving
particle}.  The thick solid black curves in Fig. \ref{FIG2}b show the mean
hydrolysis level, determined by MC simulation, in the particle
frame. The particle position is arbitrary, but the hydrolysis level
near it differs significantly from the uniform bulk away from the
particle (or in the laboratory frame). As we follow the
stochastically driven particle, the mean hydrolysis level
$\langle\sigma_{i} \rangle$ just before (after) it is higher (lower);
the particle statistically moves ahead of a domain wall, spending more
time ahead of it.


The mean velocity and dispersion of  the driven particle are computed from 

\begin{equation}
\begin{array}{l}
V = k_{+}f_{+}Q_{100} - k_{-}f_{-}Q_{001} \quad \mbox{and} \\[13pt]
D = k_{+}f_{+}Q_{100}+k_{-}f_{-}Q_{001},
\label{VV}
\end{array}
\end{equation}


\noindent where $Q_{100}$ and $Q_{001}$ are the steady-state
probabilities that the segment of three sites centered about the
driven particle is in the indicated configuration
(cf. Fig. \ref{STEPS}). In order to use mean field theory to compute
$Q_{100}$ and $Q_{001}$, we must use either moment equations or a
finite-sized mean field transition matrix in the moving, fluctuating
frame of the transported particle.  In analogy to Eq. \ref{MOMENT1},
we can consider the evolution equation of the first moment
$\langle\s_{\ell}\rangle$ of the hydrolysis level at the site of the
driven particle.  In addition to the state transitions represented by
rates $k_{\pm}$ (hydrolysis) and $k_{0}$ (dehydration), transition
terms also arise from motion of the driven particle:

\begin{equation}
\begin{array}{l}
\displaystyle {\dd \langle \s_{\ell}\rangle \over \dd t} =
k_{-}-(k_{-}+k_{0})\langle\s_{\ell}\rangle + \Delta
\langle\s_{\ell-1}\rangle - \Delta \langle\s_{\ell}\s_{\ell-1}\rangle
\\[13pt] \: \hspace{1.5cm}+\langle
p_{+}(\s_{\ell}-\s_{\ell+1})\rangle- \langle
p_{-}(\s_{\ell}-\s_{\ell-1})\rangle,
\end{array}
\label{MOMENT_V}
\end{equation}

\noindent where $p_{\pm}$ are the effective forward and backward
hopping rates of the particle. Since the particle moves only via motion of
domain walls defined by $Q_{100}$ and $Q_{001}$,

\begin{equation}
\begin{array}{rl}
p_{+} & = k_{+}f_{+}Q_{100} = k_{+}f_{+}\s_{\ell-1}(1-\s_{\ell})(1-\s_{\ell+1}) \\[13pt]
p_{-} & = k_{-}f_{-} Q_{001} = k_{-}f_{-}(1-\s_{\ell-1})(1-\s_{\ell})\s_{\ell+1}.
\end{array}
\label{P_PM}
\end{equation}

\noindent Upon substitution of $p_{\pm}$ in Eq. \ref{P_PM} into
Eq. \ref{MOMENT_V}, and neglecting all correlations, we obtain a
simple, single-site, moving-frame mean field approach to finding
$\langle\s\rangle\approx \langle\s_{\ell}\rangle \approx
\langle\s_{\ell \pm 1}\rangle$ as a root of a cubic equation.  This
simple mean field solution is only in qualitative agreement with MC
simulations. In analogy to the finite-sized segment approach
implemented through the transition matrix ${\bf M}$, we can also
improve the simple moving-frame mean field theory by considering a
sliding window of sites always centered about the driven particle
\cite{NOWAK}. Within these sites, the configurations are explicitly
enumerated, transitions involving sliding the segment as it follows
the driven particle are included, and steady-states are found.  In our
subsequent analyses, we use an $m=5$ site segment that yields
sufficiently accurate results for the parameters explored. Henceforth,
we rescale time in units of $k_{0}^{-1}$ (normalizing all rates with
respect to $k_{0}$), and set $f_{\pm}=1$.

\begin{figure}[htb]
\includegraphics[width=3.4in]{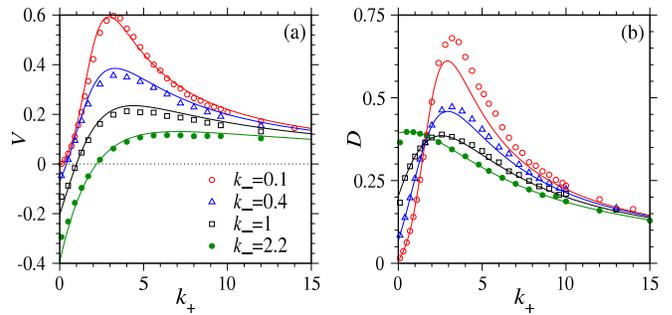}
\caption{(a) The mean velocity $V$ as a function of the forward
hydrolysis rate $k_{+}$ for various $k_{-}$.  (b) The dispersion $D$
for the same values of $k_{-}$. The symbols mark the values obtained
from MC simulation and the solid curves are results from a $5-$site
finite-segment mean field approach.}
\label{VD}
\end{figure}

Fig. \ref{VD}a shows the mean velocity $V$ derived from MC simulations
and from finite-segment mean field theory applied in the moving frame
of the convected particle. The agreement between MC simulation and the 
finite-segment MFT  is quite good provided $k_{-} \not\ll
0.1$. For $k_{-}>k_{+}$, the hydrolyzed domains grow backward at a
higher rate than forward, and the mean velocity $V<0$.  The velocity
is positive and increases once $k_{+}$ increases past
$k_{-}$. However, if $k_{+}$ becomes too large, $V$ decreases, despite
an increase in the hydrolysis asymmetry along the track.
\begin{figure}[htb]
\includegraphics[width=2.4in]{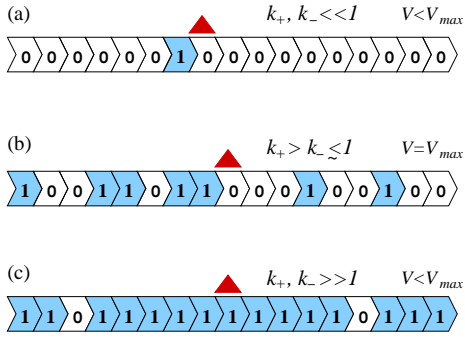}
\caption{The qualitative regimes that lead to maximum in mean velocity
$V_{max}$.  (a) If $k_{\pm}\ll 1$, most of the lattice remains
unhydrolyzed and there are few domain walls to push the particle. (b)
When $k_{+} > k_{-}$ appreciably, but not too large to destroy domain
walls, and $k_{-}\lesssim 1$, hydrolysis levels are intermediary, and
the mean velocity is maximal.  (c) For $k_{+}$ or $k_{-}$ extremely
large, the hydrolyzed domains coalesce, diminishing the number of
domain walls.  Even though the rate $k_{+}$ is large, the quantity
$k_{+}Q_{100}$ slowly diminishes.}
\label{FIG5}
\end{figure}
This behavior can be understood by considering Eq. \ref{VV} and
Fig. \ref{FIG5}.  
When both $k_{\pm} \ll 1$, dehydration dominates, domains are quickly
dissipated, and the particle is kicked by a rare 100 domain wall as
shown in Fig. \ref{FIG5}a. Increasing values of $k_{+}\gtrsim k_{-}$
modestly increases the asymmetry and hence $V$.  However, if $k_{+}$
is too large, the hydrolyzed domains merge into each other, as
depicted in Fig. \ref{FIG5}c, reducing the domain wall density
$Q_{100}$ and ultimately $V$.  Thus, there is a value of $k_{+}$ that
gives an optimally combined domain wall density and nucleation
asymmetry, as shown in Fig. \ref{FIG5}b, resulting in a maximum mean
velocity $V_{max}$.

For all values of $k_{-}$, an extremely large $k_{+}$ will ultimately
decrease the particle velocity.  Although $\Delta =k_{+}-k_{-}$ might
be large, $Q_{100}$ decreases sufficiently that $V$ decreases. The
decrease of the mean velocity $V$ in the $k_{+}\rightarrow \infty$
limit can be determined by considering a ``virial'' approximation
where only transitions among $111$, $110$, $101$, and $100$ need be
considered. Nearly all sites are hydrolyzed and we use $s\approx 1$ in
the mean field approximations. Since $100$ configurations along the
lattice are rare and spaced far apart, we can consider each triplet of
sites as independent and find $Q_{100} \approx 2(k_{0}/k_{+})^{2} +
O(k_{+}^{-3})$.  Therefore, $V \sim 2f_{+}k_{0}^{2}/k_{+}$ as
$k_{+}\rightarrow \infty$.

Fig. \ref{VD}b shows that the driven particle dispersion is maximal
near the maximum $\vert V\vert$.  This is a property of particles
with no intrinsic hopping that move stochastically only when they are
driven by domain walls.

\begin{figure}[htb]
\includegraphics[width=3.4in]{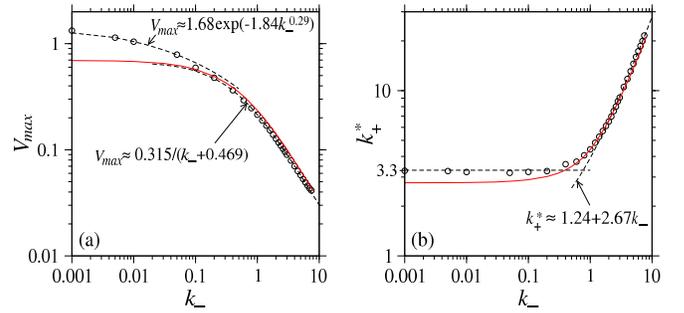}
\caption{(a) The maximum mean velocity $V_{max}$ attainable 
and (b) the value $k_{+}^{*}$ that yields this maximum
velocity for given $k_{-}$. Results derived from MC simulation
(circles) and 5-site finite-segment MFT (solid curves)
compare favorably. The dashed curves are numerical fits to the MC
simulation data given by Eqs. \ref{FIT}.}
\label{VMAX}
\end{figure}

In Fig. \ref{VMAX} we explore $V_{max}$ and the value of $k_{+}$ that
yields $V_{max}$ (designated $k_{+}^{*}$) as functions of $k_{-}$.  We
find $V_{max}$ and $k_{+}^{*}$ by accurately fitting the MC
simulations of $V$ in Fig. \ref{VD} with Pad\'{e} approximants and
indicate these optimum values in Fig. \ref{VMAX} with open
circles. For large $k_{-}$, the hydrolysis level is high and domain
merging prevails throughout the lattice (Fig. \ref{FIG5}c), requiring
ever-higher $k_{+}^{*}$ to arrive at a $V_{max}$ that is
smaller. Lowering $k_{-}$ to intermediate values increases both the
asymmetry $k_{+}-k_{-}$ and delays the onset of domain merging, giving
rise to larger $V_{max}$ at smaller values of $k_{+}^{*}$. However,
for $k_{-}\lesssim 0.1$, the hydrolyzed fraction and the onset of
domain merging become insensitive to $k_{-}$ and the optimal value of
$k_{+}$ asymptotically approaches $k_{+}^{*} \simeq 3.3$. The $m=5$
finite-segment MFT results are plotted as solid curves and for $k_{-}
> 0.1$ are in close agreement with those from MC simulations. Note
that for the case $k_{-}\rightarrow 0$ before the system size
$N\rightarrow \infty$, $\sigma_{i} = 0\, \, \forall i$ is an absorbing
but unstable condition accessible only by large deviations from the
stable steady-state that are exponentially unlikely in the system
size.  Nonetheless, for a large system a well defined steady-state can
be found and $V_{max}$ and $k_{+}^{*}$ can be accurately fit using


\begin{equation}
\begin{array}{rl}
V_{max} & \approx \displaystyle {0.315\over k_{-}+0.469} \\[13pt] 
k_{+}^{*} & \approx 1.24 + 2.67 k_{-}, \,\, k_{-} > 1, 
\label{FIT}
\end{array}
\end{equation}

\noindent and $V_{max}\approx 1.68\exp(-1.84k_{-}^{0.29})$, $k_{+}^{*}
\approx 3.3$ for $k_{-} < 0.1$.  These expressions are plotted as
dashed curves in Fig. \ref{VMAX} and provide accurate, universal
approximations on the maximum velocity possible $V_{max}$ as a
function of $k_{-}$, and the value of $k_{+}^{*}$ required for maximal
particle velocity, for each $k_{-}$.

In summary, we have presented a paradigm for substrate-driven particle
motion which has been a relatively understudied mode of subcellular
transport.  Our model captures the salient aspects of hydrolysis waves
and exhibits rich transport behavior.  Specifically, we find
short-ranged state correlations, allowing us to accurately compute
nonequilibrium steady-state particle velocities. For fixed backward
hydrolysis rate $k_{-}$, the velocities show a maximum as a function
of the forward hydrolysis rate $k_{+}$.  The value $k_{+}^{*}$
approaches its minimum near $\sim 3.3$ when $k_{-} \lesssim 0.2$. The
maximum velocities and the associated rates can be accurately
described by the simple universal fitting equations \ref{FIT}.

Additional details such as slippage ($f_{\pm} < 1$) and an external
load can be readily implemented.  A load on the particle would bias
the motion of the motor backwards and impart a negative drift velocity
in addition to that shown in Fig. \ref{VD}a. The force-velocity
relationship follows directly from the functional form of the
force-dependent drift.

\vspace{3mm}
The authors acknowledge support from the NSF (DMS-0349195)
and the NIH (K25-AI058672).

\end{document}